\documentclass[reprint,aps,nobibnotes,superscriptaddress]{revtex4-1}
\usepackage[T1]{fontenc}
\usepackage[latin9]{inputenc}
\usepackage{bm}
\usepackage{amsmath}
\usepackage{amssymb}
\usepackage{graphicx}

\usepackage{hyperref}

\newcommand{\maket}[1]{\left.\left| #1 \right\rangle\right\rangle}
\newcommand{\mabra}[1]{\left\langle\left\langle #1 \right|\right.}
\newcommand{\mainn}[2]{\left\langle\left\langle #1 \middle| #2 \right\rangle\right\rangle}
\newcommand{\ket}[1]{\left| #1 \right\rangle}
\newcommand{\bra}[1]{\left\langle #1 \right|}
\newcommand{\inn}[2]{\left\langle #1 \middle| #2 \right\rangle}
\newcommand{\Op}[3]{\left\langle #1 \middle| #2 \middle| #3 \right\rangle}
\newcommand{\vac}{\maket{\text{vac}}}
\newcommand{\cav}{\mabra{\text{vac}}}
\newcommand{\G}{\ket{\text{g}}}
\newcommand{\Kit}{r_{\rm cat}}
\newcommand{\avg}[1]{\left\langle #1 \right\rangle}
\newcommand{\Sq}[2]{\mathbb{S}_{#1}\left(#2\right)}
\newcommand{\Sqd}[2]{\mathbb{S}^{\dagger}_{#1}\left(#2\right)}
\newcommand{\ZPF}{\mathcal{M}_{\mathrm ZPF}}
\newcommand{\Tr}[1]{\mathrm{Tr}\left[#1\right]}
\newcommand{\Par}{\hat{\rho}_p}

\begin{document}

\title{Spin cat states in a ferromagnetic insulator} 
\author{Sanchar Sharma} 
\affiliation{Max Planck Institute for the Science of Light, Staudtstra\ss{}e 2, 91058 Erlangen, Germany}

\author{Victor A. S. V. Bittencourt} 
\affiliation{Max Planck Institute for the Science of Light, Staudtstra\ss{}e 2, 91058 Erlangen, Germany}

\author{Alexy D. Karenowska}
\affiliation{Clarendon Laboratory, Department of Physics, University of Oxford, Oxford OX1 3PU, United Kingdom}

\author{Silvia Viola Kusminskiy}
\affiliation{Max Planck Institute for the Science of Light, Staudtstra\ss{}e 2, 91058 Erlangen, Germany}
\affiliation{Department of Physics, University Erlangen-N{\"u}rnberg, Staudtstra\ss{}e 7, 91058 Erlangen, Germany} 

\begin{abstract}
Generating non-classical states in macroscopic systems is a long standing challenge. A promising platform in the context of this quest are novel hybrid systems based on magnetic dielectrics, where photons can couple strongly and coherently to magnetic excitations, although a non-classical state therein is yet to be observed. We propose a scheme to generate a magnetization cat state, i.e. a quantum superposition of two distinct magnetization directions, using a conventional setup of a macroscopic ferromagnet in a microwave cavity. Our scheme uses the ground state of an ellipsoid shaped magnet, which displays anisotropic quantum fluctuations akin to a squeezed vacuum. The magnetization collapses to a cat state by either a single-photon or a parity measurement of the microwave cavity state. We find that a cat state with two components separated by $\sim5\hbar$ is feasible and briefly discuss potential experimental setups that can achieve it.
\end{abstract}

\maketitle 
\emph{Introduction:} The concept of superposition is a cornerstone of quantum theory, a paradigmatic example of which is a `cat state' referring to, loosely speaking, a system which exists in a quantum superposition of two quasi-classical states. Besides their important historical link to Schrödinger's famous gedanken experiment, their insensitivity to particle loss noise \cite{Ofek_QEC} means that cat states find useful application as carriers of information (qubits) in quantum computation \cite{Jeong_QCCat,Ralph_QCCat,Mirrahimi_QCCat} or as sensors in quantum metrological tasks \cite{MunroMetro,RalphMetro,Huang_QuantumMetro,Knott_PracticalQMetro}. The robustness of a cat state increases with its size, i.e. how `distinct' the two quasi-classical components are. Experimental realizations of cat states include photon states at optical \cite{HuangCat,LundCat,WengerCat} and microwave \cite{VlastakisCat} frequencies with a size of up to $3$ and $100$ photons respectively, and a spin-state with size $\sim2\hbar$ composed of $\sim3000$ atoms \cite{McConellCat}. At a fundamental level, a macroscopic cat state is a prototypical system to study collapse in quantum-to-classical transitions \cite{Carlesso_Collapse,Stefan_Collapse}. However, non-classical states are notoriously difficult to generate in macroscopic systems due to the lack of long enough coherence lengths. Considerable advances on this front have been obtained in optomechanical systems (in which light couples to acoustic excitations \cite{COMech_Rev}) \cite{Hong17,Riedinger18,Felix20,Delic20}, however cat states have not yet been realized in this platform \cite{Shomroni_OMechCat,Zhan_OMechCat}. 

\begin{figure}[!ht]
\includegraphics[width=1\columnwidth]{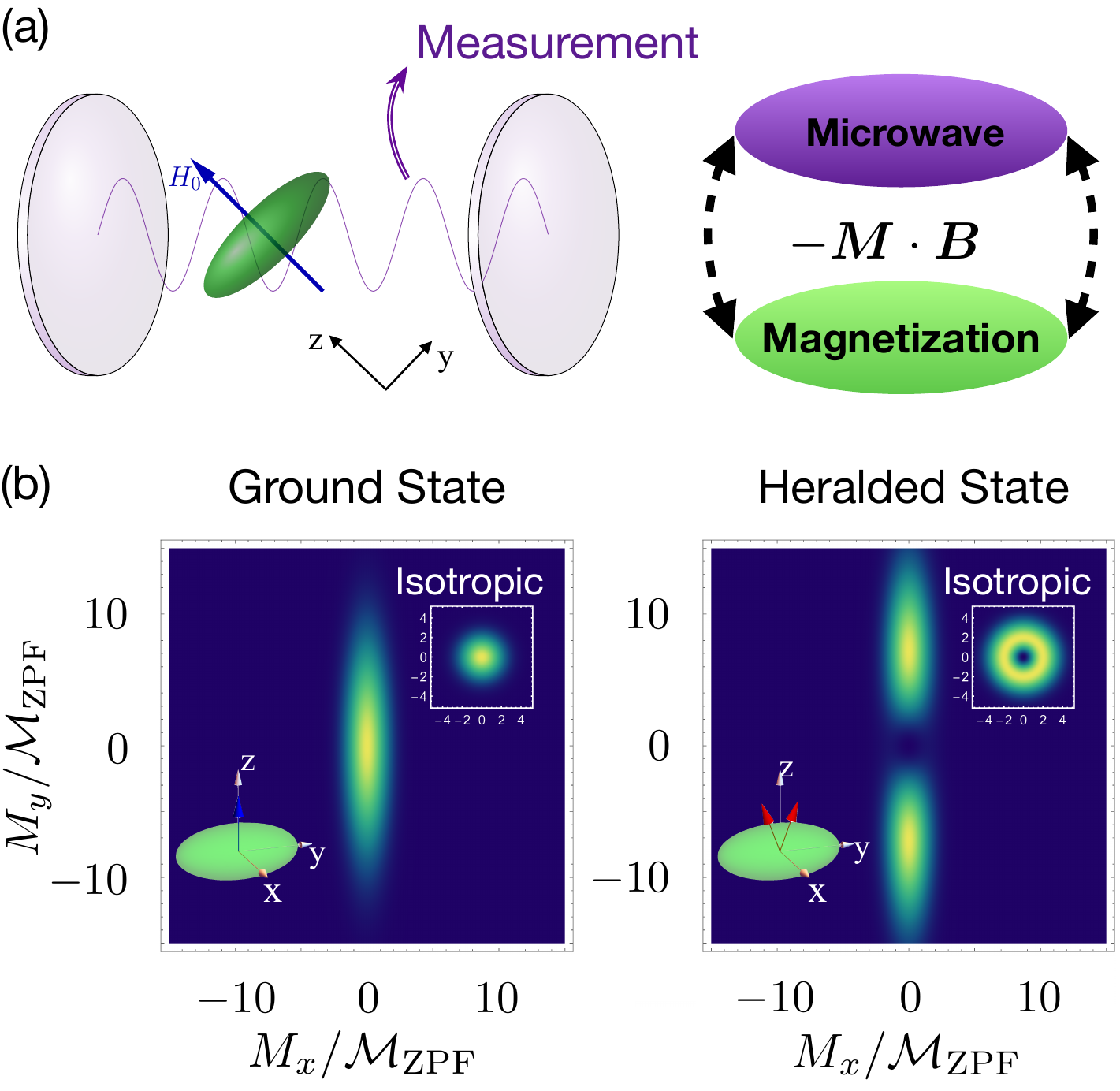}
\caption{(a) Setup: A ferromagnetic ellipsoid (green) with magnetization \textbf{$\boldsymbol{M}$} couples to microwaves in a cavity via dipole coupling to the cavity magnetic field $\bm{B}$. An external applied field $H_{0}$ saturates the magnetization along one of the short axes ($\parallel\boldsymbol{z}$).  (b) Probability density of the quasi-classical magnetization's state as a function of its components in units of the isotropic zero point fluctuations $\ZPF,$ see Eqs. ((\ref{Def:ZPF}), \ref{Def:Husimi}).  Ground state: The squeezed magnetization vacuum [see Eq. (\ref{eq:ProbGround})] with anisotropic fluctuations. The inset shows the non-squeezed (isotropic) case, valid for a spherical magnet. Heralded state: The magnetization state after detecting a microwave photon [see Eq. (\ref{eq:ProbCat})] showing features of a cat state. Such features are absent when there is no squeezing (inset).\label{fig:Setup}}
\end{figure}

Over the last few years, a new kind of hybrid quantum system has emerged as a promising platform for quantum applications, where photons are coupled coherently to magnetic excitations (magnons) in magnetic materials \cite{Lachance_Hybrid}. The dielectric ferrimagnet Yttrium Iron Garnet (YIG) is the material of choice in current experiments, owing partly to its extremely low magnetic dissipation \cite{SagaOfYIG}. Coherent and strong magnon-microwave coupling using sub-mm spheres of YIG has been realized \cite{SoykalPRL10,Zhang14,TabuchiHybrid14} and used to mediate the coupling between magnons and superconducting qubits \cite{Tabuchi_QMag}. Magnons in YIG can also coherently couple to optical photons \cite{James_OMag,Osada_OMag,Zhang_OMag,Silvia_OMag,SoykalPRL10}, phonons \cite{Zhang16,Carlos_MagPh}, and electrons \cite{Heinrich_SPCoupling,Li_SPCoherence,Woltersdorf_SPBackflow}, pointing to the possibility of magnon-based quantum transducers \cite{Ryusuke_MtoO,Zhu_MtoO}. Using YIG thin films, a Bose-Einstein condensate of magnons showing macroscopic coherence has been moreover demonstrated \cite{DemokritovBEC,RezendeBEC}. These developments together with the recent demonstration of single-magnon detection in YIG spheres \cite{Lachance17,Lachance_Magnon}, has opened prospects for studying and manipulating microwave magnetic excitations in a quantum coherent manner. Creating non-classical states of the magnetization is crucial for future applications in what has been denominated `quantum magnonics' \cite{Lachance_Hybrid}. Theoretical proposals so far include all-optical heralding of magnon Fock states \cite{Victor_Heralding} and generation of entangled states \cite{Mehrdad_NCMags,MagMWPhEnt,Li_EntYIG}.  

Here, we propose a scheme to prepare a cat state of a macroscopic number of spins ($>{10}^{18}$) which can be achieved by employing state-of-the-art microwave cavities with an embedded magnetic element of anisotropic shape, see Fig. \ref{fig:Setup}(a). The protocol relies on the anisotropy of the magnet enforcing a magnetic ground state analogous to the squeezed vacuum in quantum optics, plus the concomitant entangled spin-photon ground state when the magnet is coupled to the cavity. We show that in a YIG sample cats with a size $\sim5\hbar$ are feasible, where the size can be tuned by an external magnetic field.

\emph{Model:} A well established protocol to generate cat states in quantum optics is to add a photon to a squeezed optical vacuum \cite{DaknaCat,HuangCat,Ourjoumtsev83}. In order to accomplish this analogously in a ferromagnet, we first require a squeezed magnetization state \cite{Akash_ShotNoise}, i.e. a minimum uncertainty state with anisotropic zero-point fluctuations. This can be realized via the ground state of a magnet with an anisotropic shape, such as an ellipsoid. Notably, the degree of squeezing of the magnetic ground state can be tuned by an external magnetic field. The second step is to add an excitation, i.e. to `flip' on average one spin which is delocalized in space, see Fig. \ref{fig:Setup}(b). We show that this can be achieved by coupling the magnetization to a microwave cavity \cite{SoykalPRL10,Zhang14,TabuchiHybrid14} and performing a measurement of the latter. For low enough temperatures either a single-photon measurement \cite{Johnson_QNDSing,Nogues_QNDSing} or a parity measurement can be employed \cite{Sun2014}. We discuss these steps in detail below. 

The proposed setup is shown in Fig. \ref{fig:Setup}(a) where a ferromagnetic ellipsoid is kept inside a microwave cavity. The magnet is assumed to be slender and prolate, i.e. $L_y\gg L_x = L_z$ where $L_i$ is the length in $i$-th direction. In the absence of external magnetic fields, the magnetization would align with the longest axis. A sufficiently large field, here $H_0\boldsymbol{z}$, aligns the magnetization to $\boldsymbol{z}$ with zero-point fluctuations largely along $\boldsymbol{y}$ \cite{Akash_ShotNoise}, see Fig. \ref{fig:Setup}(b). For comparison, a spherical magnet would have isotropic zero point fluctuations $\ZPF$ (ignoring a small crystalline anisotropy of YIG) given by 
\begin{equation}
	 \ZPF^2 = \frac{\gamma\hbar M_s}{2V},\label{Def:ZPF} 
\end{equation} 
where $V$ is the volume of the magnet and $\gamma$ is the absolute value of the gyromagnetic ratio. Flipping a (delocalized) spin pushes the magnetization away from the origin. In the presence of shape anisotropy, the resulting state has the characteristic features of a cat state involving a superposition of two sufficiently distinct semiclassical states, see Fig. \ref{fig:Setup}(b). 

In the macrospin limit, the classical Hamiltonian density for the magnetization is 
\begin{equation}
	 \mathcal{H}_{\mathrm{mag}} = \frac{\mu_0}{2}\boldsymbol{M}\tilde{N}\boldsymbol{M} - \mu_0M_zH_0,\label{eq:HMag} 
\end{equation} 
where $\boldsymbol{M}$ is the total magnetization and $\tilde{N}$ is the demagnetization tensor, arising from the finite geometry. The magnitude $|\boldsymbol{M}| = M_s$ is a constant of motion \cite{StanPrabh} where $M_s$ is the saturation magnetization. For a spheroid, $\tilde{N}$ is diagonal with $N_x = N_z = N_T$ and $N_y = 1 - 2N_T$ \cite{Osborn,StanPrabh}. We assume a sufficiently large magnetic field, $H_0>M_s/2$, such that the classical ground state is $\boldsymbol{M} = M_s\boldsymbol{z}$. Since the energy cost of fluctuations in $M_y$ is smaller than that in $M_x$, the quantum fluctuations in $M_x$ and $M_y$ are different, leading to a squeezed vacuum \cite{Akash_ShotNoise}. We model the quantum fluctuations using the Holstein-Primakoff approximation \cite{HolPrim,StanPrabh} 
\begin{equation}
	 \frac{M_x - iM_y}{2\ZPF}\rightarrow\hat{s},\label{Def:HP} 
\end{equation} 
valid for $\left|M_{x,y}\right|\ll M_z$ with $\ZPF$ defined in Eq. (\ref{Def:ZPF}). Using Eq. (\ref{Def:HP}) and retaining only quadratic terms in $\hat{s}$, the magnetic Hamiltonian density (\ref{eq:HMag}) integrates to (up to a constant) 
\begin{equation}
	 \frac{\hat{H}_{{\rm mag}}}{\hbar} = \omega_0\hat{s}^{\dagger}\hat{s} + \frac{\omega_s}{2}\left(\hat{s}^2 + \hat{s}^{\dagger2}\right),\label{eq:Hmag_ellipsoid} 
\end{equation} 
where 
\begin{equation}
	 \omega_s = \left(3N_T - 1\right)\frac{\gamma\mu_0M_s}{2},\ \ \omega_0 = \gamma\mu_0H_0 - \omega_s . \label{eq:freqs} 
\end{equation} 
The bosonic operator $\hat{s}$ flips a spin from $ + \boldsymbol{z}$ to $ - \boldsymbol{z}$, satisfies the canonical commutation relation $\left[\hat{s},\hat{s}^{\dagger}\right] = 1$, and annihilates the classical ground state corresponding to all spins pointing along $ - \boldsymbol{z}$ (the spins are anti-parallel to the magnetization), $\hat{s}\ket{0} = 0$. $M_z$ is found by the constraint $|\boldsymbol{M}| = M_s$. For a sphere, $N_T = 1/3$ implying $\omega_s = 0$ and hence $\hat{s}$ ($\hat{s}^{\dagger}$) is the annihilation (creation) operator of the elementary excitations of the magnet, i.e. the magnons. Below, we consider the case of a slender prolate spheroid with $N_T \approx 1/2$ where the term $\propto\hat{s}^2 + \hat{s}^{\dagger2}$ implies that the ground state is not $\ket{0}$. 

\emph{Squeezed magnetic vacuum} - The Hamiltonian (\ref{eq:Hmag_ellipsoid}) can be diagonalized to $\hat{H}_{{\rm mag}} = \hbar\omega_m\hat{m}^{\dagger}\hat{m}$ by a Bogoliubov transformation $\hat{m} = \cosh r\hat{s} + \sinh r\hat{s}^{\dagger}$, with 
\begin{equation}
	 \omega_m = \sqrt{\omega_0^2 - \omega_s^2} = \gamma\mu_0\sqrt{H_0\left(H_0 - \frac{M_s}{2}\right)} . \label{Eq:MagFreq} 
\end{equation} 
 The parameter $r$ characterizes the degree of squeezing and is given by 
\begin{equation}
	 e^r = \sqrt{\frac{\omega_0 + \omega_s}{\omega_m}} = \left(1 - \frac{M_s}{2H_0}\right)^{- 1/4} . \label{Eq:DegSq} 
\end{equation} 
The ground state of $\hat{H}_{\mathrm{mag}}$, defined by $\hat{m}\G = 0$, is given by \cite{Akash_ShotNoise,WallsQO} $\G = \Sq{r}{\hat{s}}\ket{0}$ where 
\begin{equation}
	 \Sq{r}{\hat{s}} = \exp\left[\frac{r\left(\hat{s}^2 - \hat{s}^{\dagger2}\right)}{2}\right]\label{Def:Sqzng} 
\end{equation} 
is the \emph{squeezing operator}. In the presence of anisotropy, the ground state of the system $\G$ is therefore characterized by a finite number of `flipped spins', $\Op{g}{\hat{s}^{\dagger}\hat{s}}{g} = \sinh^2r$. 

The characteristics of the ground state $\G$ can be visualized in terms of semiclassical magnetization states $\ket{\alpha}$, defined as the ones with an average magnetization such that $M_x - iM_y = 2\ZPF\alpha$ and presenting minimum fluctuations [cf. Eq. (\ref{Def:HP})]. These are given by coherent states \cite{GlauberCoherent}, satisfying $\hat{s}\ket{\alpha} = \alpha\ket{\alpha}$ defined by $\ket{\alpha} = \hat{D}(\alpha)\ket{0}$ where the displacement operator is $\hat{D}(\beta) = \exp\left[\beta\hat{s}^{\dagger} - \beta^*\hat{s}\right]$. For a general state $\ket{\psi}$, the Husimi Q-function 
\begin{equation}
	 Q(\alpha,\ket{\psi}) = \frac{1}{\pi}\left|\inn{\alpha}{\psi}\right|^2\label{Def:Husimi} 
\end{equation} 
can be interpreted as the probability density of $\ket{\psi}$ being near the semi-classical state $\ket{\alpha}$. For the ground state $\G$, defined above, we find 
\begin{equation}
	 Q(\alpha,\G) = \frac{1}{\pi\cosh r}\exp\left[ - \frac{\alpha_R^2e^r + \alpha_I^2e^{- r}}{\cosh r}\right],\label{eq:ProbGround} 
\end{equation} 
where $\alpha = \alpha_R + i\alpha_I$ with real $\alpha_{R,I}$. This is shown in Fig. \ref{fig:Setup}(a) demonstrating that fluctuations in $M_y$ are larger than that in $M_x$, indicating a squeezed vacuum. The degree of squeezing, Eq.~(\ref{Eq:DegSq}), becomes arbitrarily high as $H_0\rightarrow M_s/2$. In this limit, however, $\omega_m\rightarrow0$ and the system goes towards an instability signaling a significant change in the classical ground state and a consequent failure of the linearization used in Eq. (\ref{Def:HP}). In practice however, the frequency $\omega_m$ is bounded by an experimentally feasible low temperature. Considering $H_0$, slightly higher than $M_s/2$, such that $\omega_m = 2\pi\times100\,\text{MHz}$ (corresponding to a cryogenic temperature of $\hbar\omega_m/k_B = 5\text{mK}$) we obtain an upper limit $e^r = 5$, where we used $\gamma\mu_0M_s = 2\pi\times5\,\text{GHz}$ for YIG \cite{StanPrabh} corresponding to $\mu_0H_0 \approx 70\mathrm{kA/m}$. 

\emph{Coupling to a microwave cavity.-- }The classical Hamiltonian density for a hollow cavity reads 
\begin{equation}
	 \mathcal{H}_{\mathrm{cav}} = \frac{\epsilon_0|\boldsymbol{E}(\boldsymbol{r})|^2}{2} + \frac{|\boldsymbol{B}(\boldsymbol{r})|^2}{2\mu_0} . 
\end{equation} 
Typically, a microwave cavity hosts multiple electromagnetic modes in the GHz range. Considering a magnon frequency $\sim100\mathrm{MHz}$, the system is in the dispersive coupling regime and all of these modes have to be taken into account as long as they have good overlap with the magnet. However, for the purpose of our analysis, we can assume that the cavity consists of a single mode since the generalization does not change the qualitative features. The quantization $\boldsymbol{B}(\boldsymbol{r})\rightarrow\boldsymbol{B}_0(\boldsymbol{r})\hat{a} + \boldsymbol{B}_0^*(\boldsymbol{r})\hat{a}^{\dagger}$ where $\boldsymbol{B}_0$ is the mode profile and $\hat{a}$ is the annihilation operator of the cavity mode (and analogously for $\boldsymbol{E}(\boldsymbol{r})$), diagonalizes the cavity Hamiltonian $\hat{H}_{\mathrm{cav}} = \hbar\omega_a\hat{a}^{\dagger}\hat{a}$ up to a constant. The magnetization couples to the microwave fields via $\mathcal{H}_{\mathrm{coup}} =  - \boldsymbol{M}\cdot\boldsymbol{B}(\boldsymbol{r})$ inside the magnet and $\mathcal{H}_{\mathrm{coup}} = 0$ outside. For magnets much smaller than the microwave's wavelength $\sim\mathrm{cm}$, the magnetic field is nearly constant inside the magnet. From Eq. ((\ref{Def:HP})), we get 
\begin{equation}
	 \frac{\hat{H}_{\mathrm{coup}}}{\hbar} = \left(g_{{\rm -}}^*\hat{s}\hat{a}^{\dagger} + g_{{\rm -}}\hat{s}^{\dagger}\hat{a}\right) + \left(g_{{\rm +}}^*\hat{s}^{\dagger}\hat{a}^{\dagger} + g_{{\rm +}}\hat{s}\hat{a}\right), 
\end{equation} 
where we ignored a tertiary term $\propto\hat{s}^{\dagger}\hat{s}\hat{a}$ with a coefficient smaller than $g_{\pm}$ by a factor $\sim\ZPF/M_s$. The beam-splitter ($g_-$) and parametric-amplifier ($g_+$) coupling strengths are given by 
\begin{equation}
	 g_{\pm} =  - V\ZPF B_{0\pm}\left(\boldsymbol{r}_{\mathrm{magnet}}\right) 
\end{equation} 
where $B_{0\pm} = B_{0x}\pm iB_{0y}$ and $\boldsymbol{r}_{\mathrm{magnet}}$ is the position of the magnet inside the cavity. For simplicity, we consider circularly polarized photons with $B_{0 +} = 0$ (hence $g_+ = 0$) and define $g_-\equiv g$. By changing the global phase of photons, if necessary, we can assume $g>0$. Depending on the experimental setup, $g$ is tunable up to a large fraction of the cavity's frequency \cite{Bourhill16,Potts_Cavity}. 

We discuss now the ground state of the total Hamiltonian (see App. \ref{App:MagMW} for details)
\begin{equation}
	 \frac{\hat{H}}{\hbar} = \omega_0\hat{s}^{\dagger}\hat{s} + \frac{\omega_s}{2}\left(\hat{s}^2 + \hat{s}^{\dagger2}\right) + \omega_a\hat{a}^{\dagger}\hat{a} + g\left(\hat{s}\hat{a}^{\dagger} + \hat{s}^{\dagger}\hat{a}\right). \label{eq:Hamiltonian} 
\end{equation} 
We assume $\omega_m,g\ll\omega_s,\omega_a$ implying a large spin squeezing [cf. Eqs. (\ref{Eq:MagFreq},\ref{Eq:DegSq})]. The Hamiltonian has two eigenfrequencies $\{\Omega_a,\Omega_m\}$ where $\Omega_a \approx \omega_a$ and 
\begin{equation}
	 \Omega_m \approx \sqrt{\omega_m^2 - \frac{2g^2\omega_0}{\omega_a}}\label{Eq:Om} 
\end{equation} 
is dispersively shifted from the bare magnon's frequency $\omega_m$. For large couplings, $g>\omega_a(\omega_0 - \omega_s)$, the system becomes unstable. 

The ground state of the spin-photon system is given by $\vac = \Sq{R_m}{\hat{w}_m}\Sq{R_a}{\hat{w}_a}\maket{0}$ with $\hat{w}_m = \hat{s}\cos\theta/2 + \hat{a}\sin\theta/2$, and $\hat{w}_a =  - \hat{s}\sin\theta/2 + \hat{a}\cos\theta/2$. It therefore consists of a non-zero number of photons and spin-flips given by a joint squeezed operation over the classical ground state $\maket{0}$ defined by $\hat{s}\maket{0} = \hat{a}\maket{0} = 0$. The general expressions for $\{R_m,R_a,\theta\}$ are given in App. \ref{App:MagMW}. For $\omega_m,g\ll\omega_s,\omega_a$ we get 
\begin{equation}
	 \theta \approx \frac{- 2g}{\omega_a}, 
\end{equation} 
implying that $\hat{w}_m$ and $\hat{w}_a$ have respectively a large spin and photon contribution with small mixing, characteristic of an off-resonant coupling. In this case $R_a \approx 0$, and we obtain 
\begin{equation}
	 e^{R_m} \approx \sqrt{\frac{2\omega_s}{\Omega_m}} . \label{Eq:Dm} 
\end{equation} 
As the coupling $g$ increases, $\Omega_m$ decreases by the dispersive shift [see Eq. (\ref{Eq:Om})], and thus the squeezing parameter $R_m$ increases. For $g = 0$, $\Omega_m = \omega_m$ and $e^{R_m} \approx e^r$ [see Eq. (\ref{Eq:DegSq})]. 

In the ground state$\vac$, assuming $\omega_m,g\ll\omega_s,\omega_a$, the average number of spin-flips is 
\begin{equation}
	 \avg{\hat{s}^{\dagger}\hat{s}} \approx \frac{e^{2R_m}}{4} \approx \frac{\omega_s}{2\Omega_m}, 
\end{equation} 
while the average photon number is 
\begin{equation}
	 \avg{\hat{a}^{\dagger}\hat{a}} \approx \frac{g^2\avg{\hat{s}^{\dagger}\hat{s}}}{\omega_a^2},\label{eq:navac} 
\end{equation} 
where the averages are taken w.r.t $\vac$. Eq. (\ref{eq:navac}) is expected from Fermi's Golden rule, valid for weak coupling. In the limit of infinite squeezing, $\Omega_m\rightarrow0$, the number of spin flips (and, correspondingly, of photons) diverges. For moderate squeezing (e.g. $e^{R_m}\sim5$ as taken below considering temperature constraints) and $g\ll\omega_a$, the number of photons in $\vac$ is, however, very small. 

\begin{figure}
\includegraphics[width=1\columnwidth]{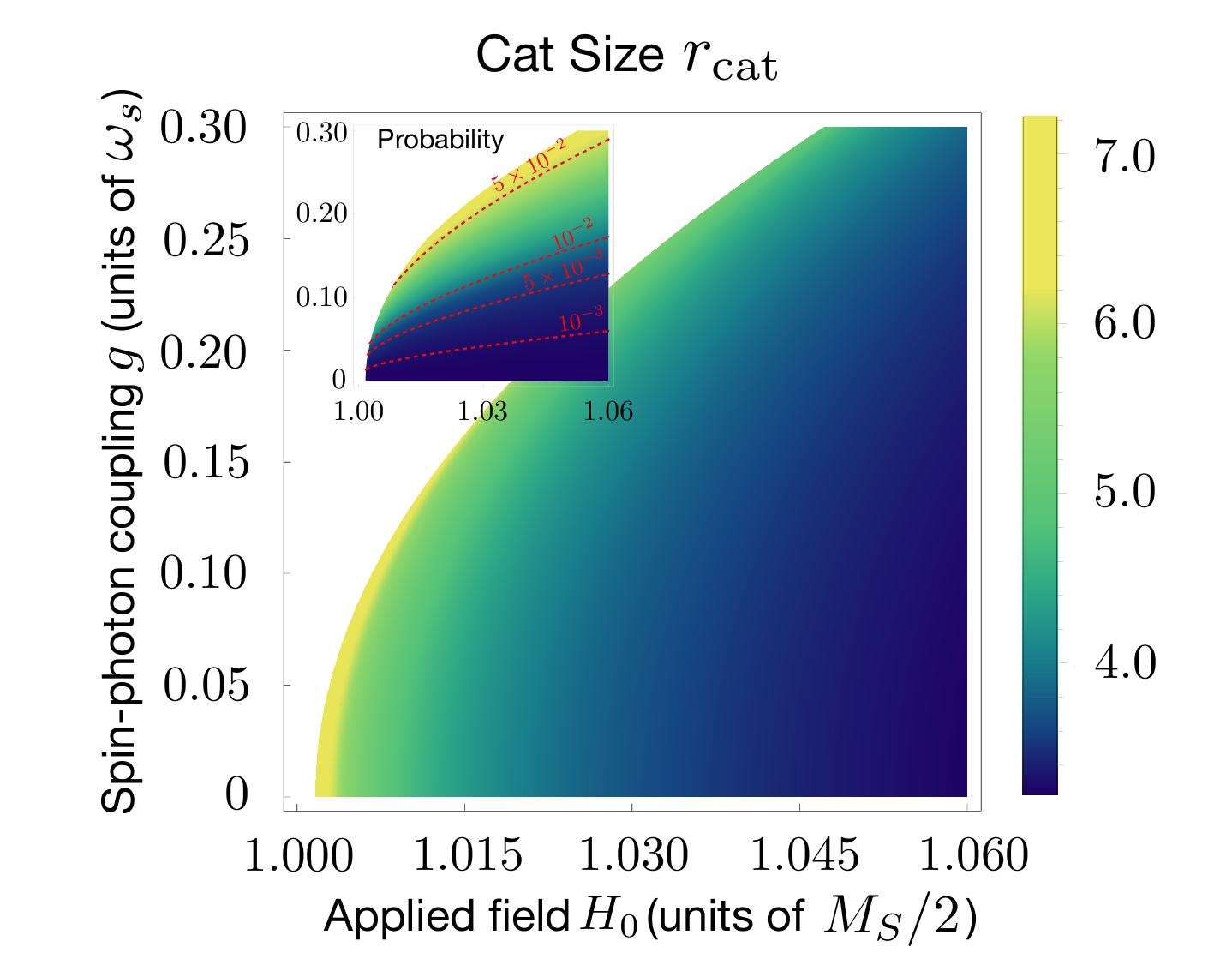}
\caption{Cat size $r_{\text{cat}}$ and probability (inset, logarithmic scale) of finding the cavity in a single-photon state $P$ as a function of applied field $H_{0}$ and spin-photon coupling $g$ for $\omega_{s}=\omega_{a}=2\pi\times5\mathrm{GHz}$ (valid for YIG \cite{StanPrabh}). The white region contains the parameter space where either the system is unstable or the magnon's frequency is too low $\Omega_{m}<2\pi\times100\mathrm{MHz}$.\label{fig:Res}}
\end{figure}

\emph{Results:} We now show how the coupling of the magnet to a microwave cavity can be used to herald a cat state of the magnetization. The joint spin-photon ground state $\vac$ is an entangled state, and thus a measurement of the state of the cavity can affect the magnetization in a non-trivial way. In particular, for a measurement projecting the ground state $\vac$ to a single-photon state, we find that the state of the magnetization collapses to [see App. \ref{App:SinglePh}] 
\begin{equation}
	 \ket{C} = \frac{1}{\cosh R}\hat{s}^{\dagger}\ \Sq{R}{\hat{s}}\ket{0},\label{eq:CState} 
\end{equation} 
which corresponds to flipping a spin from the magnetization's squeezed vacuum $\ket{g} = \Sq{R}{\hat{s}}\ket{0}$ with $R$ given by 
\begin{equation}
	 \tanh R = \frac{\omega_0 - \Omega_m + \omega_a - \Omega_a}{\omega_s} . 
\end{equation} 
In the general case, we can interpret $R$ as the effective magnetization squeezing, as opposed to $R_m$ which is the squeezing of the hybridized mode $\hat{w}_m$. For $\omega_m,g\ll\omega_s,\omega_a$, $\hat{w}_m \approx \hat{s}$ and consequently, $e^R \approx e^{R_m}$ . In this regime, the main control parameters are the effective magnetization squeezing $R$, tunable via the external magnetic field, and spin-photon coupling $g$, tunable by magnet's position. 

To understand the properties of the magnetization state $\ket{C}$, we consider the probability density defined by the Husimi Q-function Eq. (\ref{Def:Husimi}). We obtain 
\begin{equation}
	 Q(\alpha,\ket{C}) = \frac{\alpha_R^2 + \alpha_I^2}{\pi\cosh^3R}\exp\left[ - \frac{\alpha_R^2e^r + \alpha_I^2e^{- r}}{\cosh R}\right],\label{eq:ProbCat} 
\end{equation} 
for $\alpha = \alpha_R + i\alpha_I$ with real $\alpha_{R,I}$. This is plotted in Fig. \ref{fig:Setup}(b) showing two regions of high probability. Specifically, we can separate the upper and lower lobes, $\ket{C}\propto\ket{C_+} - \ket{C_-}$ where 
\begin{equation}
	 \ket{C_{\pm}}\propto\sum_r\frac{(\pm1)^r\sqrt{r!}}{\Gamma(r/2 + 1/2)}\left(\frac{- \tanh R}{2}\right)^{\frac{r - 1}{2}}\ket{r} . 
\end{equation} 
Explicit calculations show 
\begin{equation}
	 \inn{C_-}{C_+} \approx \frac{8e^{- 3R}}{3\pi}, 
\end{equation} 
where we ignored terms higher order in $e^{- R}$, therefore the two components are nearly orthogonal for $e^R\sim5$. $Q(\alpha,\ket{C})$ has two peaks at $\alpha = \pm i\Kit/2$ with the cat size 
\begin{equation}
	 \Kit = \sqrt{2\left(e^{2R} + 1\right)} \approx 2\sqrt{\frac{\omega_s}{\Omega_m}},\label{eq:Size} 
\end{equation} 
where the approximation holds for $\omega_s,\omega_a\gg\Omega_m\gg g^2/\omega_a$ \footnote{We note that the cat size seems to be non-zero when $R \approx 0$, but this is merely an artifact of the definition which is irrelevant for $\Kit\sim5$.}.  The probability of finding the cavity in a single-photon state is given in terms of states with $1$-photon and $n$-spin flips, $\maket{1,n}\propto\hat{a}^{\dagger}\left(\hat{s}^{\dagger}\right)^n\maket{0}$, as 
\begin{equation}
	 P = \sum_n\left|\mainn{1,n}{\text{vac}}\right|^2 . \label{eq:P_exact} 
\end{equation} 
For $\omega_s,\omega_a\gg\Omega_m\gg g^2/\omega_a$ we find [see Eq. (\ref{eq:ProbEx})] $P \approx \avg{\hat{a}^{\dagger}\hat{a}}$, where the average number of photons is given in Eq. (\ref{eq:navac}). This expression for the probability is expected, since in this limit the average number of cavity photons is small [see below Eq. (\ref{eq:navac})]. 

The system will be in the ground state if $k_BT<\hbar\Omega_m$, which, as discussed for $\omega_m$, effectively imposes an experimental lower limit on $\Omega_m$. We consider therefore an applied field such that the lower hybridized mode has the frequency $\Omega_m = 2\pi\times100\,\text{MHz}$, giving an upper temperature limit of $T<5\,\text{mK}$. Taking as an example $\omega_a = \omega_s$ and $g = 0 . 05\omega_a$, we obtain a squeezing parameter $e^R = 5$, a cat size $\Kit = 7$, and a heralding probability $P = 0 . 03$. 

In Fig.~\ref{fig:Res}, we plot the cat size $r_{\text{cat}}$ and heralding probability $P$ as a function of external magnetic field $H_0$ and spin-photon coupling $g$. The plots are generated using the exact expressions given in App. \ref{App:SinglePh}, instead of the approximate ones discussed here. They show a trade-off: the maximum achievable probability increases with increasing $g$, as expected, while the cat size decreases. This is the case since a larger coupling $g$ puts a lower limit on the magnon's frequency [cf. Eq. (\ref{Eq:Om})] and consequently an upper limit on the squeezing of the magnetization [cf. Eq. (\ref{Eq:DegSq})]. 

Single photon detection in microwave cavities typically involves long protocols and significant errors \cite{Johnson_QNDSing,Nogues_QNDSing}. However, in the limit of small photon numbers, zero and one photon states can also be distinguished by their parity, which can be measured with a high accuracy by coupling the cavity to a qubit \cite{Sun2014}. Projecting onto the odd-parity photon state, the density matrix of the spins is given by a partial trace over the photons 
\begin{equation}
	 \Par = \mathrm{Tr}_{\hat{a}}\left[\frac{I - ( - 1)^{\hat{a}^{\dagger}\hat{a}}}{2}\vac\cav\right] . 
\end{equation} 
As shown in App. \ref{App:Parity}, the probability of finding the cavity in an odd-parity state in this limit is aprroximatley the one-photon probability $P$. In order to compare $\Par$ with $\ket{C}$, we use the fidelity measure \cite{Jozsa_fidelity,Uhlmann_fidelity} $F = \Op{C}{\Par}{C}$ [see App. \ref{App:Parity}] which can be interpreted as the probability of finding the magnetization in the state $\ket{C}$ (\ref{eq:CState}). When $\omega_s,\omega_a\gg\Omega_m\gg g^2/\omega_a$, one can show 
\begin{equation}
	 1 - F\sim\frac{2P^2}{3} . 
\end{equation} 
As $P<0 . 1$ [see Fig. \ref{fig:Res}], we get a very high fidelity $F>0 . 99$, implying that our results can be used with a parity measurement as well. 

\emph{Experimental considerations:} For the design of practical experiments to create the cat state, a range of different approaches can be envisaged. As discussed in detail, our proposal is reliant on the creation of a squeezed state of the magnetization of the YIG system achieved through a pronounced anisotropy. Thus far, we have assumed that this anisotropy is geometric although other sources of anisotropy (for example, crystalline anisotropy) could be exploited to the same end. Nevertheless, we choose to focus on shape anisotropy both for consistency and because it likely represents the most practical route to performing an experiment using the current state-of-the-art in measurement technology.

The maximum dimension of the YIG sample employed is set by the maximum length over which coherence can be maintained which, in high-purity monocrystalline YIG, is expected to be over $100 \mathrm{\mu m}$ \cite{Andrich17,Lachance17,Lachance_Magnon}. In order to perform a photon measurement, the YIG sample must be installed in a microwave cavity coupled to at least one Josephson junction-based qubit. Though a range of different cavity and qubit styles may be envisaged for this purpose, the main decision to be made is whether to employ a planar or 3D geometry. A 3D geometry has the benefit of lending itself to a spatial separation of the field-sensitive qubit and the small bias field required to saturate the magnetic sample. Moreover, such systems have already been used to demonstrate a range of important results in the context of quantum measurements on magnon systems \cite{Lachance17,Lachance_Magnon}. For the purposes of the present measurement, however, such an arrangement has the disadvantage that, since each dimension of the cavity must be at least a substantial fraction of the vacuum microwave wavelength at a few GHz ($\lambda\sim\mathrm{cm}$), the volume of the microwave mode will be very large in comparison to the volume of the YIG sample, placing a fundamental limit on the achievable coupling (we require $\sim100\mathrm{MHz}$, see Fig. \ref{fig:Res}). Conversely, with a planar (quasi one-dimensional) geometry, while the challenge of confining the bias field requires a more creative solution, the system has the advantage that the microwave mode is strongly confined to a volume that can be as much as $6$ orders of magnitude smaller than in the 3D case ($\lambda d^2\sim1\mathrm{cm}\times1\mu\mathrm{m}\times1\mu\mathrm{m}\sim10^{- 6}\lambda^3$ where $d$ is the resonator width). Accordingly, we suggest that an elegant way to measure the cat state would be to use the now classic methodology first proposed by Schuster et al. \cite{Schuster07}. A relatively simple microwave quantum circuit could be constructed in which the YIG ellipsoid sits in the dielectric gap of a planar superconducting resonator coupled to a judiciously positioned transmon qubit. Spectroscopy would be performed on the system and the herald photon number measured via the occupancy-dependent Stark shift of the qubit.  

\emph{Conclusions:} We proposed a scheme to generate a magnetization cat state in a spin-microwave hybrid system. The scheme relies on adding a quanta to a squeezed vacuum of magnetization that is realized as the ground state of an anisotropic magnet \cite{Akash_ShotNoise}. We showed that cat states with the two components differing by $\Kit\hbar\sim5\hbar$ can be generated in sub-mm YIG samples, comfortably within the precision range of current quantum measurements of magnetization \cite{Lachance17,Lachance_Magnon}. The size of the cat state is larger when the cavity is measured to be in higher photon numbers [see App. (\ref{App:SinglePh})], although the heralding probabilities are much smaller. The lifetime of the cat states is given by inverse of the magnon's linewidth $\sim{10}^4/\omega_m\sim0 . 1\text{ms}$ \cite{Kingler_Gilbert}. Our analysis is valid when the system is in its ground state giving experimentally feasible temperature restrictions $T<5\text{mK}$. We envision our results to expand the field of quantum magnonics and applications of ferromagnets as quantum transducers and ultrasensitive magnetic field sensors, and to pave the way for protocols involving truly non-classical macroscopic states of magnetization. 

\emph{Acknowledgements:} We thank J. Haigh for valuable discussion. SS, VASVB, and SVK acknowledge financial support from the Max Planck Society.  

\appendix 

\section{Ground state in a general Bogoliubov Transformation} \label{App:BogT}

In this section, we review the multi-mode Bogoliubov transformation \cite{Derezinski_BQH,Colpa_BQH} which we use in the main text to operate on the spin-microwave coupled system, and find the ground state of a quadratic Hamiltonian assuming stability, i.e. that the eigenvalues of the Hamiltonian are lower bounded. We treat the general case of $N$-modes, as the $2$-mode case is not much simpler, and the former can be generalized to the case when all cavity modes are included and multiple magnets are present in the cavity. 

Consider a set of harmonic oscillators with annihilation operators $\hat{u}_1,\dots,\hat{u}_N$. We define column vectors $\hat{U} = \left(\hat{u}_1,\dots,\hat{u}_N\right)^T$ and $\hat{U}^* = \left(\hat{u}_1^{\dagger},\dots,\hat{u}_N^{\dagger}\right)^T$ along with their row counterparts $\hat{U}^T$ and $\hat{U}^{\dagger} = (\hat{U}^*)^T$. We have the canonical commutation relations $\left[\hat{u}_i,\hat{u}_j\right] = 0$, and $\left[\hat{u}_i,\hat{u}_j^{\dagger}\right] = \delta_{ij}$ where $\delta_{ij}$ is the Kronecker delta. 

Any Hermitian quadratic Hamiltonian can be written as 
\begin{equation}
	 \hat{H} = \hbar\hat{U}^{\dagger}A\hat{U} + \frac{\hbar}{2}\hat{U}^{\dagger}B\hat{U}^* + h . c . ,\label{GenHamMat} 
\end{equation} 
where $A$ and $B$ are $N\times N$ matrices. We can choose $A = A^{\dagger}$ and $B = B^T$. 

A Bogoliubov transformation is performed by defining a new set of independent harmonic oscillators $\hat{v}_1,\dots,\hat{v}_N$ 
\begin{equation}
	 \hat{V} = D^{\dagger}\hat{U} + N^{\dagger}\hat{U}^* . \label{GenBogo} 
\end{equation} 
As $\hat{v}_i$ must satisfy canonical commutation relations, we have the constraint 
\begin{equation}
	 D^{\dagger}D - N^{\dagger}N = I,\ \ N^TD = D^TN . \label{Conds:UV1} 
\end{equation} 
These conditions imply the inverse transformation 
\begin{equation}
	 \hat{U} = D\hat{V} - N^*\hat{V}^* . 
\end{equation} 
Again, using the fact that both $\hat{u}_i$ and $\hat{v}_i$ satisfy canonical commutation relations, we find 
\begin{equation}
	 DD^{\dagger} - N^*N^T = I,\ \ ND^{\dagger} = D^*N^T . \label{Conds:UV2} 
\end{equation} 
Below, see Eq.~(\ref{GenForm:UV}), we find the most general form of $D$ and $N$ satisfying these constraints. 

The matrices $U$ and $V$ are found by diagonalizing 
\begin{equation}
	 \tilde{H} = \begin{pmatrix}
	A & - B\\
	 B^* & - A^* 
\end{pmatrix} . \label{GenHJ} 
\end{equation} 
It can be shown that the eigenvalues of $\tilde{H}$ are real when the Hamiltonian $\hat{H}$ is diagonalizable with positive frequencies \cite{Derezinski_BQH,Colpa_BQH}. For each eigenvalue $\Omega>0$, let the eigenvector be $\left(d^T\ n^T\right)$ where $d$ and $n$ are vectors of size $N$. Then, we can normalize $|d|^2 - |n|^2 = 1$. Furthermore, for each eigenlist $\{\Omega,d,n\}$ we also have $\{- \Omega,d^*,n^*\}$ \cite{Derezinski_BQH,Colpa_BQH}. Thus, we can write $\tilde{H}T = T\tilde{\Omega}$ where $\tilde{\Omega}$ is a diagonal matrix with entries $\{\Omega_1,\dots,\Omega_N, - \Omega_1,\dots, - \Omega_N\}$ and the diagonalization matrix 
\begin{equation}
	 T = \begin{pmatrix}
	D & N^*\\
	 N & D^* 
\end{pmatrix} 
\end{equation} 
defines $D$ and $N$. With this transformation, it can be shown that the Hamiltonian (\ref{GenHamMat}) is diagonalized so that one can write, up to a constant, $\hat{H} = \sum_i\hbar\Omega_i\hat{v}_i^{\dagger}\hat{v}_i$ . 

The ground state of $\hat{H}$, $\vac$, is given by the state that is annihilated by all $\hat{v}_i$, which in terms of $\hat{u}_i$ is 
\begin{equation}
	 \left(D^{\dagger}\hat{U} + N^{\dagger}\hat{U}^*\right)\vac = 0 . 
\end{equation} 
To find this state in terms of the excitations in the original $\hat{U}$-basis, we write $D$ and $N$ in Bloch-Messiah form \cite{Braunstein05}, 
\begin{equation}
	 N ={\cal S}^*\sinh R{\cal S}^{\dagger}{\cal F},\ \ D ={\cal S}\cosh R{\cal S}^{\dagger}{\cal F}, 
\end{equation} 
where ${\cal S}$ and ${\cal F}$ are unitary and $R$ is a diagonal matrix with positive entries. This is the most general form where all the identities Eq.~(\ref{Conds:UV1},\ref{Conds:UV2}) are satisfied. For completeness, we give an elementary proof of this below [see near and after Eq. (\ref{eq:Proof1})]. 

The ground state can be written in a simplified form in terms of the set of harmonic oscillators defined through the unitary operation $\hat{W} ={\cal S}^{\dagger}\hat{U}$. Then, the ground state is given by a set of single-particle identities 
\begin{equation}
	 \left(\cosh R_i\hat{w}_i + \sinh R_i\hat{w}_i^{\dagger}\right)\vac = 0 . 
\end{equation} 
The above expression can be simplified via the squeezing operators $\Sq{R_i}{\hat{w}_i}$ [see the definition (\ref{Def:Sqzng})], which are unitary and satisfy \cite{WallsQO} 
\begin{equation}
	 \Sq{R_i}{\hat{w}_i}\,\hat{w}_i\,\Sqd{R_i}{\hat{w}_i} = \cosh R_i\hat{w}_i + \sinh R_i\hat{w}_i^{\dagger} . 
\end{equation} 
The ground state is $\vac = \mathbb{S}\ket{0}$ where the `bare vacuum' is defined by $\hat{u}_i\maket{0} = 0$ and the squeezing operator is 
\begin{equation}
	 \mathbb{S} = \prod\Sq{R_i}{\hat{w}_i} . \label{Def:SqTot} 
\end{equation} 
The quadratures $\hat{w}_i + \hat{w}_i^{\dagger}$, which are a linear combination of the quadratures of $\hat{u}_i$, are squeezed by $e^{R_i}$. 

In the remaining section, we show how to find ${\cal S}$, ${\cal F}$, and $R$ in terms of $D$ and $N$. The Hermitian matrix $DD^{\dagger}$ has all eigenvalues $>1$ because $DD^{\dagger} = I + N^*N^T$ [Eq.~(\ref{Conds:UV2})]. It can be diagonalized as $DD^{\dagger} = \tilde{{\cal S}}\cosh^2R\tilde{{\cal S}}^{\dagger}$ which implies $NN^{\dagger} = \tilde{{\cal S}}^*\sinh^2R\tilde{{\cal S}}^T$. 

Consider the symmetric matrix $N_s = \tilde{{\cal S}}^*\sinh R\tilde{{\cal S}}^{\dagger}$, which satisfies $N_sN_s^{\dagger} = NN^{\dagger}$. This implies $N = N_s\tilde{F}_N$ where $\tilde{F}_N$ is unitary. Then, we also have the diagonalization, 
\begin{equation}
	 N^{\dagger}N = \tilde{{\cal F}}_N^{\dagger}\tilde{{\cal S}}\sinh^2R\tilde{{\cal S}}^{\dagger}\tilde{{\cal F}}_N . \label{eq:Proof1} 
\end{equation} 
Similarly, consider the Hermitian matrix $D_H = \tilde{{\cal S}}\cosh R\tilde{{\cal S}}^{\dagger}$ which implies $D = D_H\tilde{{\cal F}}_D$ with unitary $\tilde{{\cal F}}_D$. Then, 
\begin{equation}
	 D^{\dagger}D = \tilde{{\cal F}}_D^{\dagger}\tilde{{\cal S}}\cosh^2R\tilde{{\cal S}}^{\dagger}\tilde{{\cal F}}_D . 
\end{equation} 
Using $D^{\dagger}D - N^{\dagger}N = I$, 
\begin{equation}
	 \tilde{{\cal F}}_D^{\dagger}\tilde{{\cal S}}\cosh^2R\tilde{{\cal S}}^{\dagger}\tilde{{\cal F}}_D = \tilde{{\cal F}}_N^{\dagger}\tilde{{\cal S}}\cosh^2R\tilde{{\cal S}}^{\dagger}\tilde{{\cal F}}_N 
\end{equation} 
implies that $\tilde{{\cal D}}_{{\rm ph}} = \tilde{{\cal S}}^{\dagger}\tilde{{\cal F}}_N\tilde{{\cal F}}_D^{\dagger}\tilde{{\cal S}}$ commutes with the diagonal matrix $\cosh^2R$. If all the diagonal entries of $R$ are distinct, $\tilde{{\cal D}}_{{\rm ph}}$ is diagonal. As it is also unitary, it consists of phases as the diagonal entries. These phases can be traced back to the freedom in choosing the phases of the eigenvectors of $DD^{\dagger}$, i.e. choosing the phases of each column of $\tilde{{\cal S}}$. When $R$ has some degeneracies, we also have freedom to choose linear combinations of the degenerate eigenvectors. . The absence of a full set of constraints complicates the proof but the result can be recovered by assuming small perturbations and a limiting process. 

Now, we redefine the phases in the eigenmatrix ${\cal S} = \tilde{{\cal S}}\sqrt{\tilde{{\cal D}}_{{\rm ph}}^*}$ and redo the exercise above. We find ${\cal F}_D = \tilde{{\cal F}}_D$ and ${\cal F}_N = \tilde{{\cal S}}\tilde{\mathcal{D}}_{{\rm ph}}^*\tilde{{\cal S}}^{\dagger}\tilde{{\cal F}}_N$. This implies ${\cal F}_N ={\cal F}_D ={\cal F}$. Summarizing, 
\begin{equation}
	 N ={\cal S}^*\sinh R{\cal S}^{\dagger}{\cal F},\ \ D ={\cal S}\cosh R{\cal S}^{\dagger}{\cal F} . \label{GenForm:UV} 
\end{equation} 
For analytical purposes, it is easier to look at the decomposition (known as Takagi factorization), 
\begin{equation}
	 ND^{- 1} ={\cal S}^*\tanh R{\cal S}^{\dagger} . \label{GenVUinv} 
\end{equation} 
This becomes the eigenvalue decomposition when ${\cal S}$ is real and the eigenvalues of $ND^{- 1}$ are positive. 

\section{Coupled spin microwave system\label{App:MagMW}} 

In this section, we apply the general theory described in App. \ref{App:BogT} to the special case of coupled spin-microwave system. We solve first for the ground state exactly and later simplify it to the case of high spin squeezing. 

The Hamiltonian is given by Eq.~(\ref{GenHamMat}) with $\hat{U} = \left(\hat{s},\hat{a}\right)^T$ and the matrices 
\begin{equation}
	 A = \begin{pmatrix}
	\omega_0 & g\\
	 g & \omega_a 
\end{pmatrix},\ \ B = \begin{pmatrix}
	\omega_s & 0\\ 0 & 0 
\end{pmatrix} . 
\end{equation} 
The eigenfrequencies found by diagonalizing Eq.~(\ref{GenHJ}) are $\{\Omega_m,\Omega_a\}$, where 
\begin{equation}
	 \Omega_m^2 + \Omega_a^2 = 2g^2 + \omega_m^2 + \omega_a^2, 
\end{equation} 
and 
\begin{equation}
	 \Omega_m\Omega_a = \sqrt{g^4 - 2g^2\omega_a\omega_0 + \omega_a^2\omega_m^2}, 
\end{equation} 
with bare magnon frequency $\omega_m = \sqrt{\omega_0^2 - \omega_s^2}$. Both $\Omega_m,\Omega_a>0$ iff $g<\sqrt{\omega_a(\omega_0 - \omega_s)}$, while the system is unstable for higher $g$. Using the eigenvectors, the Bogoliubov transformation defined in Eq.~(\ref{GenBogo}) is given by 
\begin{equation}
	 D = \omega_s
\begin{pmatrix}
	\mathcal{N}_a(\Omega_a^2 - \omega_a^2) & \mathcal{N}_m(\Omega_m^2 - \omega_a^2)\\
	 \mathcal{N}_ag(\Omega_a + \omega_a) & \mathcal{N}_mg(\Omega_m - \omega_a) 
\end{pmatrix} 
\end{equation} 
and 
\begin{equation}
	 N = \begin{pmatrix}
 - \mathcal{N}_a\Sigma_a(\Omega_a + \omega_a) & - \mathcal{N}_a\Sigma_a(\Omega_a + \omega_a)\\
	 \mathcal{N}_a\Sigma_ag & \mathcal{N}_a\Sigma_ag 
\end{pmatrix}, 
\end{equation} 
where $\Sigma_{m,a} = \left(\Omega_{m,a} - \omega_0\right)\left(\Omega_{m,a} - \omega_a\right) - g^2$ and the normalization constants $\mathcal{N}_{m,a}$ can be found by enforcing $D^TD - N^TN = I$ [see Eq.~(\ref{Conds:UV1})]. 

The directions and degrees of squeezing can be found using Eq.~(\ref{GenVUinv}). We have 
\begin{equation}
	 ND^{- 1} = \begin{pmatrix}
	S_m & S_T\\
	 S_T & S_a 
\end{pmatrix} 
\end{equation} 
where 
\begin{align}
	 S_m &= \frac{\omega_0 + \omega_a - \Omega_+ - \Omega_-}{\omega_s}\label{Def:Sm}\\
	 S_T &= \frac{g^2 + (\omega_a - \Omega_+)(\omega_a - \Omega_-)}{\omega_sg}\label{Def:ST}\\ S_a &=  - g\frac{(\omega_a + \omega_0)S_T - gS_m}{(\omega_a + \Omega_+)(\omega_a + \Omega_-)}\label{Def:Sa} 
\end{align} 
Note that $ND^{- 1}$ is symmetric as expected from Eq.~(\ref{Conds:UV1}). Then, the ground state is given by $\vac = \Sq{R_m}{\hat{w}_m}\Sq{R_a}{\hat{w}_a}\maket{0}$ where $\hat{w}_m = \hat{s}\cos\theta/2 + \hat{a}\sin\theta/2$ and $\hat{w}_a =  - \hat{s}\sin\theta/2 + \hat{a}\cos\theta/2$ with mixing parameter 
\begin{equation}
	 \tan\theta = \frac{2S_T}{S_m - S_a},\label{Def:theta} 
\end{equation} 
and the squeezing parameters 
\begin{align}
	 \tanh R_m &= S_m + S_T\tan\frac{\theta}{2},\\
	 \tanh R_a &= S_a - S_T\tan\frac{\theta}{2} . 
\end{align}

Consider the case when $\omega_0 \approx \omega_s$, i.e. $\omega_m\ll\omega_s,\omega_a$. For stability, we require a small $g<\sqrt{\omega_a(\omega_0 - \omega_s)}$. Then, $\Omega_a \approx \omega_a$ and 
\begin{equation}
	 \Omega_m \approx \sqrt{\omega_m^2 - \frac{2g^2\omega_0}{\omega_a}} . 
\end{equation} 
The squeezing directions are given by 
\begin{equation}
	 \theta \approx \frac{- 2g}{\omega_a} . 
\end{equation} 
The degrees of squeezing are $R_a \approx 0$ and $e^{R_m} \approx \sqrt{2\omega_s/\Omega_m}$. As $|\theta|\ll1$, there is a strong squeezing in a quadrature close to $S_x\propto\hat{s} + \hat{s}^{\dagger}$. 

\section{Single Photon Detection} \label{App:SinglePh}

In App. \ref{App:MagMW}, we derived the multimode squeezed ground  state. When the cavity is measured to be in a single photon Fock state, we expect that the magnetization collapses to a cat state. We derive the size of the cat state here. 

The ground state of the cavity-spin coupled system  is given by squeezing the position of the harmonic oscillators $\hat{w}_m = \hat{s}\cos\theta/2 + \hat{a}\sin\theta/2$ and $\hat{w}_a =  - \hat{s}\sin\theta/2 + \hat{a}\cos\theta/2$. Explicitly, 
\begin{multline}
	\vac=\frac{1}{\sqrt{\cosh R_{m}\cosh R_{a}}}\\
		\exp\left[-\frac{\tanh R_{m}\hat{w}_{m}^{\dagger,2}+\tanh R_{a}\hat{w}_{a}^{\dagger,2}}{2}\right]\maket{0}.\label{eq:VacApp}
\end{multline}

Here $\{\theta,R_m,R_a\}$ are defined in App. \ref{App:MagMW}. We Taylor expand the above in $\hat{a}^{\dagger}$, 
\begin{equation}
	 \vac = \sum_n\sqrt{P_n}\hat{{\cal O}}_n\frac{\hat{a}^{\dagger,n}}{\sqrt{n!}}\maket{0} . 
\end{equation} 
$\hat{{\cal O}}_n$ operates only on the spin subspace and satisfies $\Op{0}{\hat{\mathcal{O}}_n^{\dagger}\hat{\mathcal{O}}_n}{0} = 1$. $P_n$ is the probability of finding the ground state $\vac$in an $n$-photon Fock state. 

Projecting the ground state onto a $1$-photon Fock state, the magnetization collapses to $\ket{C} = \hat{{\cal O}}_1\ket{0}$ where (by explicit Taylor expansion) 
\begin{equation}
	 \hat{{\cal O}}_1 = \frac{1}{\cosh^{3/2}R}\hat{s}^{\dagger}\exp\left[ - \frac{\tanh R\ \hat{s}^{\dagger,2}}{2}\right]\label{MagOp1} 
\end{equation} 
with $\tanh R = S_m$ [see Eq. \ref{Def:Sm}]. This corresponds to flipping a spin over a squeezed vacuum with degree $R$. Correspondingly, the probability is given by 
\begin{equation}
	 P_1 = S_T^2\frac{\cosh^3R}{\cosh R_m\cosh R_a} . \label{eq:ProbEx} 
\end{equation} 
The magnetization state can be expressed in a basis of semi-classical states, introduced before Eq. (\ref{Def:Husimi}) and given by 
\begin{equation}
	 \ket{\beta} = e^{- |\beta|^2/2}\sum_{n = 0}^{\infty}\frac{\beta^n}{\sqrt{n!}}\ket{n}, 
\end{equation} 
as 
\begin{equation}
	 \inn{\beta}{C} = \frac{1}{\cosh^{3/2}R}\beta^*\exp\left[ - \frac{\left|\beta\right|^2 + \beta^{*,2}\tanh R}{2}\right] 
\end{equation} 
For a fixed $|\beta|$, $\left|\inn{\beta}{C}\right|$ is maximized at $\beta = \pm i|\beta|$. By differentiating w.r.t. $|\beta|$, we find the maxima at $\beta = \pm i\Kit/2$ with 
\begin{equation}
	 \Kit = \sqrt{2\left(e^{2R} + 1\right)} . 
\end{equation}

The size of the cat state is larger for a projective measurement of $n>1$ photons, which we briefly discuss here. Given the Wigner function of the joint spin-photon vacuum (\ref{eq:WTot}) [to be discussed in detail in App. \ref{App:Parity}], the state of the magnetization after a projection into a $n$-photon Fock state \cite{CarlosIntro} 
\begin{equation}
	 W_{{\rm AM}}(\alpha_s,\hat{U},\hat{\rho}_g) = 4\pi\int d^2\alpha_aW(\alpha_s,\alpha_a,\hat{U},\hat{\rho}_g)W_n(\alpha_a), 
\end{equation} 
where the integration is performed over the photon variables and 
\begin{equation}
	 W_n(\alpha_a) = \frac{( - 1)^n}{2\pi}L_n(4\vert\alpha_a\vert^2)\,e^{- 2\vert\alpha_a\vert^2}, 
\end{equation} 
is the Wigner function for the $n$ Fock state, with $L_n$ the $n^{{\rm th}}$ order Laguerre polynomial. 

\begin{figure}
\centering{}\includegraphics[width=1\columnwidth]{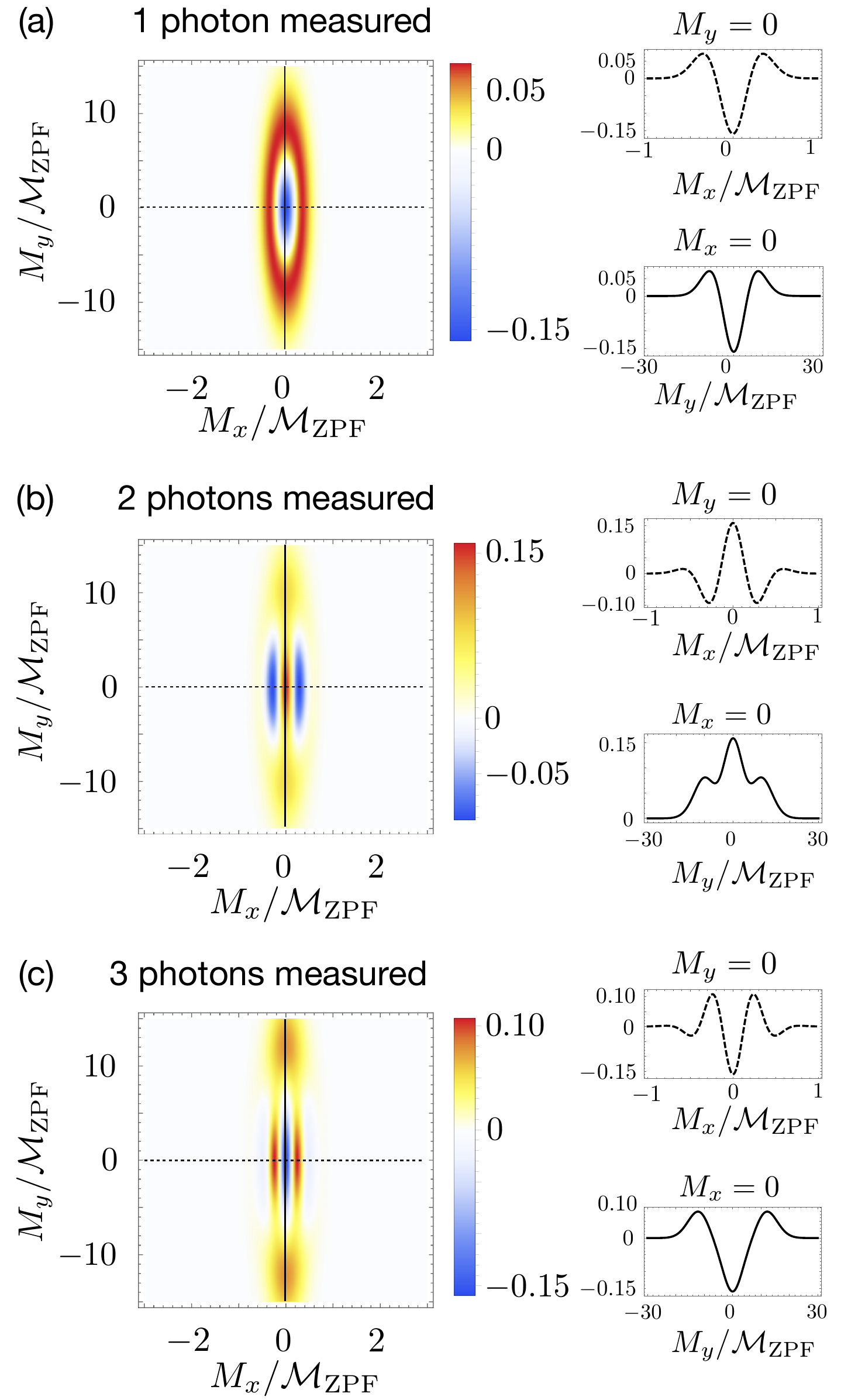}
\caption{Wigner function of the heralded state after the measurement of ${1,2,3}$ photon(s). The Wigner function resembles that of a squeezed cat state, with the characteristic negative part and interference pattern close to the origin. Parameters: $\omega_{a}=\omega_{s}$, $g=0.05\omega_{a}$ and applied field adjusted such that $\Omega_{m}=\omega_{a}/12.5$.} \label{FigApp:WigFunc}
\end{figure}

In Fig. \ref{FigApp:WigFunc} we show the Wigner function of the heralded state after one photon measured [cf. Eq. (\ref{Wig:SingPh})]. Its profile resembles, as expected, a squeezed cat state with minor fringes (two maxima and one minimum) near the origin. By measuring more than one photon, the heralded states have larger cat sizes with the downside of smaller heralding probabilities. 

\section{Parity Detection\label{App:Parity}} 

In App. \ref{App:SinglePh}, we showed that detecting the cavity in a single-photon state heralds the magnetization into a cat-like state. However, as the expected average number of photons is small, we expect similar results for detecting the cavity in an odd parity state given by the projection, 
\begin{equation}
	 \hat{\Pi}_a = \frac{\hat{I} - ( - 1)^{\hat{a}^{\dagger}\hat{a}}}{2}, 
\end{equation} 
which is experimentally less demanding. To find the magnetization's heralded state, we use the Wigner function defined as the Fourier transform 
\begin{equation}
	 W(\alpha,\hat{C},\hat{\rho}) = \int\frac{d^{2N}\alpha}{\pi^4}e^{\beta^{\dagger}\alpha - \alpha^{\dagger}\beta}\chi(\beta,\hat{C},\hat{\rho}),\label{Def:Wig} 
\end{equation} 
where ${\alpha,\beta,\hat{C}}$ are $2\times1$ column vectors, $\hat{\rho}$ is a density matrix, and the characteristic function $\chi(\beta,\hat{C},\hat{\rho})$ is defined as 
\begin{equation}
	 \chi(\beta,\hat{C},\hat{\rho}) = \Tr{\hat{\rho}D(\beta,\hat{C})}\label{Def:chi} 
\end{equation} 
with the multi-mode displacement operator 
\begin{equation}
	 D(\beta,\hat{C}) = \exp\left[\hat{C}^{\dagger}\beta - \beta^{\dagger}\hat{C}\right] . \label{Def:Disp} 
\end{equation} 
As discussed in Sec. \ref{App:MagMW}, the ground state is $\vac = \Sq{R_m}{\hat{w}_m}\Sq{R_a}{\hat{w}_a}\maket{0}$ in terms of the harmonic oscillator pair $\hat{W} = (\hat{w}_m\ \hat{w}_a)^T$ defined by $\hat{W} ={\cal S}^T\hat{U}$ where $\hat{U} = (\hat{s}\ \hat{a})^T$ and 
\begin{equation}
	{\cal S} = \begin{pmatrix}
	\cos\theta/2 & - \sin\theta/2\\
	 \sin\theta/2 & \cos\theta/2 
\end{pmatrix} 
\end{equation} 
with $\theta$ defined in Eq. (\ref{Def:theta}). For the ground state $\hat{\rho}_g = \vac\cav$, we find a Gaussian Wigner function, 
\begin{equation}
	 W(\alpha,\hat{W},\hat{\rho}_g) = \left(\frac{2}{\pi}\right)^2\exp\left[ - 2\left(\alpha_R^Te^{2R}\alpha_R + \alpha_I^Te^{- 2R}\alpha_I\right)\right],\label{eq:WGenW} 
\end{equation} 
where $\alpha_{R,I}$ are real vectors satisfying $\alpha = \alpha_R + i\alpha_I$ and $R$ is the diagonal matrix with entries $\{R_m,R_a\}$. This can be converted to original basis using 
\begin{equation}
	 W(\alpha,\hat{U},\hat{\rho}) = W(\mathcal{S}^{\dagger}\alpha,\hat{W},\hat{\rho}),\label{eq:WGenU} 
\end{equation} 
giving 
\begin{equation}
	 W(\alpha,\hat{U},\hat{\rho}_g) = \left(\frac{2}{\pi}\right)^2\exp\left[ - 2\left(\alpha_R^T\Sigma_-^{- 1}\alpha_R + \alpha_I^T\Sigma_+^{- 1}\alpha_I\right)\right],\label{eq:WTot} 
\end{equation} 
where variance matrices are 
\begin{equation*}
	 \Sigma_{\pm} = \mathcal{S}e^{\pm2R}\mathcal{S}^T . 
\end{equation*}

In the following, we suppress the harmonic oscillator indices, e.g. the ground state characteristic function is $\chi_g(\{\alpha_s,\alpha_a\})$ and Wigner function is $W_g(\{\alpha_s,\alpha_a\})$, given in Eq. (\ref{eq:WTot}). Similar to Eq. (\ref{Def:chi}), we can also define the characteristic function of the magnetization alone 
\begin{equation}
	 \chi_{s,g}(\alpha_s) = \Tr{\hat{\rho}_gD(\alpha_s,\hat{s})}, 
\end{equation} 
and then, the Wigner function is obtained by tracing out the photons, 
\begin{equation}
	 W_{s,g}(\alpha_s) = \int d^2\alpha_aW_g(\{\alpha_s,\alpha_a\}) . 
\end{equation} 
This integrates to 
\begin{equation}
	 W_{s,g}(\alpha_{sR} + i\alpha_{sI}) = \frac{2}{\pi\mathcal{N}_{s,g}}\exp\left[ - 2\left(\frac{\alpha_{sR}^2}{\sigma_-^2} + \frac{\alpha_{sI}^2}{\sigma_+^2}\right)\right],\label{eq:Wsg} 
\end{equation} 
where the variances are 
\begin{equation}
	 \sigma_{\pm}^2 = e^{\pm2R_1}\cos^2\frac{\theta}{2} + e^{\pm2R_2}\sin^2\frac{\theta}{2}, 
\end{equation} 
and the normalization constant is 
\begin{equation}
	 \mathcal{N}_{s,g} = \sigma_+\sigma_- = \sqrt{1 + \sin^2\theta\sinh^2(R_1 - R_2)} . 
\end{equation} 
After measurement, the composite state is given by the density matrix 
\begin{equation}
	 \hat{\rho}_{{\rm AM}} = \frac{\hat{\Pi}_a\hat{\rho}_g\hat{\Pi}_a}{P_p}, 
\end{equation} 
where $P_p$ is the probability of getting odd parity found by enforcing $\Tr{\hat{\rho}_{\mathrm{AM}}} = 1$ or equivalently the normalization, 
\begin{equation}
	 \int d^2\alpha_sd^2\alpha_aW_{\mathrm{AM}}(\{\alpha_s,\alpha_a\}) = 1,\label{Norm:2HO} 
\end{equation} 
To find the magnetization's Wigner function, consider its characteristic function [again suppressing other arguments in Eq. (\ref{Def:chi})], 
\begin{equation}
	 \chi_{s,\mathrm{AM}}(\alpha_s) = \frac{1}{P_p}\Tr{\hat{\Pi}_a\hat{\rho}_g\hat{\Pi}_aD(\alpha_s,\hat{s})} . 
\end{equation} 
Using $\hat{\Pi}_a^2 = \hat{\Pi}_a$ and its expansion in displacement operators \cite{Royer77}, 
\begin{equation}
	 2\hat{\Pi}_a = \hat{I} - \int\frac{d^2\alpha_a}{2\pi}D(\alpha,\hat{a}), 
\end{equation} 
we find 
\begin{equation}
	 2P_p\chi_{s,\mathrm{AM}}(\alpha_s) = \chi_{s,g}(\alpha_s) - \int\frac{d^2\alpha_a}{2\pi}\chi_g(\{\alpha_s,\alpha_a\}) . 
\end{equation} 
Taking the Fourier transform, we get the heralded Wigner function of the magnetization 
\begin{equation}
	 2P_pW_{s,\mathrm{AM}}(\alpha_s) = W_{s,g}(\alpha_s) - \frac{\pi}{2}W_g(\{\alpha_s,0\}), 
\end{equation} 
which can be expanded using Eq. (\ref{eq:Wsg}) and 
\begin{equation}
	 W_g(\{\alpha_s,0\}) = \left(\frac{2}{\pi}\right)^2\exp\left[ - 2\left(\sigma_+^2\alpha_{sR}^2 + \sigma_-^2\alpha_{sI}^2\right)\right] . 
\end{equation} 
The normalization of Wigner function, Eq. (\ref{Norm:2HO}), gives 
\begin{equation}
	 P_p = \frac{1}{2} - \frac{1}{2\mathcal{N}_{s,g}} . 
\end{equation} 
For small $|\theta|e^{2R_m},R_a\ll1$, we get 
\begin{equation}
	 P_p \approx \frac{\theta^2e^{2R_m}}{16}, 
\end{equation} 
same as that of single photon detection [see below Eq. (\ref{eq:P_exact})], as expected because of small photon numbers.  

To compare the parity-heralded state, say $\hat{\rho}_p$, with the 1-photon-heralded state $\hat{\rho}_1 = \ket{C}\bra{C}$, we consider the fidelity measure 
\begin{equation}
	 F = \Tr{\sqrt{\sqrt{\hat{\rho}_1}\hat{\rho}_p\sqrt{\hat{\rho}_1}}}^2 . 
\end{equation} 
It is easy to show $\sqrt{\hat{\rho}_1} = \hat{\rho}_1$ giving $F = \Tr{\hat{\rho}_1\hat{\rho}_p}$. This can be written in terms of Wigner functions as \cite{QOSchleich} 
\begin{equation}
	 \Tr{\hat{\rho}_1\hat{\rho}_p} = \pi\int d^2\alpha W_1(\alpha)W_p(\alpha) 
\end{equation} 
where $W_p\equiv W_{s,\mathrm{AM}}$ and (by explicit calculations) 
\begin{equation}
	 W_1(\alpha) = \frac{8}{\pi}\left[\alpha_R^2e^{2R} + \alpha_I^2e^{- 2R} - \frac{1}{4}\right]e^{- 2\left(\alpha_R^2e^{2R} + \alpha_I^2e^{- 2R}\right)} . \label{Wig:SingPh} 
\end{equation} 
Integrating, we get 
\begin{equation}
	 F = \frac{4\mathcal{N}_{s,g}\left(1 + \mathcal{N}_{s,g}\right)}{\left(e^{- 2R} + \sigma_-^2\right)^{3/2}\left(e^{2R} + \sigma_+^2\right)^{3/2}} 
\end{equation} 
For $g^2/\omega_a\ll\Omega_m,g\ll\omega_s,\omega_a$, 
\begin{equation}
	 1 - F\sim\frac{g^4\omega_s^2}{6\omega_a^4\Omega_m^2} \approx \frac{2P_p^2}{3} . 
\end{equation} 
Typically, the probability of heralding is $<0 . 1$ [see Fig. \ref{fig:Res}], so the fidelity is very high implying that the two state $\hat{\rho}_1$ and $\hat{\rho}_p$ are almost same. 

%


\end{document}